\documentclass[showpacs, pra,twocolumn,preprintnumbers ,amsmath, amssymb, superscriptaddress, aps]{revtex4-2}
\usepackage{amsfonts}
\usepackage{color}
\usepackage{amsmath,amssymb}
\usepackage{pifont}
\usepackage{amssymb}  
\usepackage{bbold}
\usepackage{float}
\usepackage{subfloat}

\usepackage[caption=false]{subfig}
\usepackage{tikz}
\usepackage{makecell}
\usepackage{subfig}
\usepackage{pifont}   
\usepackage{graphicx} 
\graphicspath{{Figures4/}}
\usepackage{dcolumn}  
\usepackage{bm}       
\usepackage{multirow} 
\usepackage{placeins}
\usepackage[colorlinks]{hyperref}
\usepackage{mathtools}
\usepackage{appendix}

\def \be{\begin{align}}
	\def \ee{\end{align}}
\def \bea{\begin{eqnarray}}
	\def \eea{\end{eqnarray}}

\begin{document}

        \title{
        Transmissions and group delay time in graphene with proximity exchange field and double barriers }
        \date{\today}
     \author{Ahmed Jellal}
        \email{a.jellal@ucd.ac.ma}
        \affiliation{Laboratory of Theoretical Physics, Faculty of Sciences, Choua\"ib Doukkali University, PO Box 20, 24000 El Jadida, Morocco}
       \author{Rachid El Aitouni}
               \affiliation{Laboratory of Theoretical Physics, Faculty of Sciences, Choua\"ib Doukkali University, PO Box 20, 24000 El Jadida, Morocco} 
               
               	\author{Pablo Díaz}
               \affiliation{Departamento de Ciencias F\'{i}sicas, Universidad de La Frontera, Casilla 54-D, Temuco 4811230, Chile}  
            		\author{David Laroze}
            \affiliation{Instituto de Alta Investigación, Universidad de Tarapacá, Casilla 7D, Arica, Chile}

        
\begin{abstract} 
	
We study the transmission and group delay time for fermions in graphene under a proximity exchange field scattered by double barriers. Solving the Dirac equation over five regions, we calculate transmission and reflection coefficients using the transfer matrix method, and analyze group delay time using a Gaussian wave packet and the stationary phase method. Our results reveal spin-dependent features in transmission and group delay time, with notable shifts between spin orientations, especially for configurations with up to three layers of boron nitride (BN). We observe enhanced Klein tunneling peaks and full transmission conditions for certain combinations of system parameters. The double-barrier configuration also significantly improves the group delay time compared to the single-barrier case. In fact, we show that the group delay time oscillates as the barrier width increases without showing signs of saturation, indicating the absence of the Hartman effect. This is in contrast to the single-barrier case, where the group delay time is found to saturate as the barrier width increases. In addition, we identify critical angles and maximum energies for evanescent modes.


\end{abstract}          
        
\maketitle

\section{Introduction}

The transport of electrons in graphene is {modeled} by a Dirac-like equation, which allows the study of relativistic quantum phenomena \cite{Geim, Alhaidari}. The electronic states form perfect Dirac cones when energies are close to the degeneracy point \cite{Malko}. One of the wonderful things about Dirac fermions in graphene is their ability to tunnel across high and wide potential barriers with unit probability, which is called Klein tunneling or the Klein paradox \cite{Calogeracos, Stander, Giavaras2009}, in contrast to the ordinary tunneling of non-relativistic particles. Obviously, the electrons can pass through a potential barrier greater than their energy to reach the perfect transmission, contrary to the traditional non-relativistic tunneling \cite{Bai, Pereira}. The secret of quantum tunneling is the possibility for the quantum particles to pass into forbidden regions as allowed by Heisenberg's uncertainty principle \cite{Allain}.

In recent decades, {quantum tunneling \cite{Yampolskii2008, Bliokh2010,Rozhkov2011}.} has become more prominent, especially with respect to the time it takes a particle to tunnel through a potential barrier \cite{Martin, Hauge}. Since quantum tunneling is a phenomenon dealing with superluminal effects, the group delay time has also turned out to be an important quantity in explaining the dynamical aspects of the process. For example, Hartman showed that for a particle tunneling through a rectangular barrier, the group delay time becomes constant regardless of the thickness of the barrier, as long as the barrier is opaque \cite{Hartman}. This result has been called the Hartman effect \cite{Hartman, Nimtz}. It essentially states that for sufficiently large barriers, the effective group velocity \cite{Olkhovsky} of the particle can exceed the speed of light, which has been proposed and demonstrated in experiments with optical systems \cite{Chiao}. First, Enders and Nimtz proved the existence of the Hartman effect under laboratory conditions with a waveguide in which a short section acted as a barrier to waves of frequencies below the cutoff frequency of that region \cite{Enders, Enders93}. 
These studies have provided the basis for further studies of wave propagation in quantum systems. More recently, graphene with various barrier nanostructures has been the focus of several studies, especially those involving tunneling time measurements. In these studies, time is treated as a parameter rather than an observable in the framework of quantum mechanics  \cite{177,Fattasse2022}.
A recent extension of the conventional transfer matrix method has been proposed to account for anisotropic features in electron transmission in two-dimensional materials \cite{Bautista2024}. This approach offers a valuable mathematical tool for studying the diverse properties of such materials.

In the absence of a gap, the Dirac fermions lack a finite mass, breaking the equivalence between the two carbon sublattices of graphene and making the use of graphene in electronic devices very challenging. To induce a bandgap in the electronic spectrum of graphene, hexagonal boron nitride ($h$-BN) is an effective option. When exposed to an external potential, graphene opens a band gap at the Dirac point \cite{San}. This wide-gap insulator has a similar structure to graphene but introduces important differences. Indeed, when graphene is placed on $h$-BN, the interaction with the substrate breaks the equivalence of its two carbon sublattices and shifts the Fermi level into the gap, transforming graphene into a semiconductor \cite{Giovannetti, Moon}.
The ability of electrons to propagate over long distances in graphene results in a weak spin-orbit coupling, which is of particular interest for opening a gap at the Dirac points on the order of $10^{-3}$ meV \cite{Jahani, Yao}. To enhance this spin-orbit coupling, the magnetic proximity effect of an adjacent ferromagnetic insulator (FMI) is exploited \cite{Liu, Haugen}. In this context, cobalt (Co) and nickel (Ni) are introduced into graphene via a few (one to three) layers of hexagonal boron nitride ($h$-BN) \cite{Zollner, Tepper}. Graphene provides an exceptional platform for the study of proximity-induced phenomena, making it an outstanding material in the family of 2D materials \cite{Singh}. 
Recently, the effect of magnetic proximity on valley filtering in graphene has been studied in \cite{Zheng2022}. In addition, various trends in trilayer graphene have been explored, including the effects of proximity spin-orbit and exchange couplings in ABA and ABC configurations \cite{Zollner2022}. The influence of these couplings on the correlated phase diagram of rhombohedral trilayer graphene has also been studied \cite{Zhumagulov2024}. This range of research enhances our understanding of how the electronic properties of graphene-based materials can vary in response to different external fields due to proximity-induced interactions.

The electron tunneling in graphene-based junctions has been studied in \cite{Tepper}. Graphene was assumed to be either directly on a ferromagnetic insulator or separated from a metallic ferromagnetic substrate by a few atomic layers of boron nitride (BN).  The tunneling was found to be spin dependent, and the group delay time was shown to exhibit the Hartman effect. We extend these results to the double rectangular barrier scenario. We solve the Dirac equation over five different regions to derive the energy spectrum of the whole system. We then calculate the transmission and reflection coefficients using the transfer matrix method. The study of the group delay time is performed by analyzing a Gaussian wave packet, with the transmitted and reflected phase times derived by the stationary phase method. We discuss the numerical results and study the tunneling and group delay time as a function of incident angle, barrier height, barrier width, and energy of the incoming electrons for both spin-up and spin-down states.
Our results indicate that for the double-barrier case, both transmission and group delay time exhibit spin-dependent characteristics, with noticeable shifts between the spin-up and spin-down states, especially evident for up to three layers of BN. As the double-barrier structure becomes more sophisticated, Klein tunneling peaks become more pronounced, and full transmission occurs for certain combinations of incident angle, barrier width, and incident electron energy. In addition, the group delay time increases significantly. In particular, we show that the Hartman effect is absent, as the group delay time fluctuates with increasing barrier width and shows no signs of saturation.
  Finally, we recover the results from the single-barrier model in \cite{Tepper} by setting \(v = u = V\) and \(d = D/2\), demonstrating that our generalized formulation builds on and enriches the previous results.

The paper is organized as follows. In Sec. \ref{II}, we begin by formulating the problem, presenting the Hamiltonian describing the system, and deriving the corresponding energy spectrum solution. In Sec. \ref{III}, we apply boundary conditions to construct the transfer matrix and use it to calculate the reflection and transmission probabilities. Sec. \ref{IV} is devoted to the computation of group delay time, using stationary states to determine these quantities. In Sec. \ref{V}, we perform a detailed numerical analysis of our results, followed by a discussion. Finally, we summarize and conclude our work in Sec. \ref{VI}.

\section{Theoretical model}\label{II}
\begin{figure}[h] 
	\centering
	\includegraphics[scale=0.9]{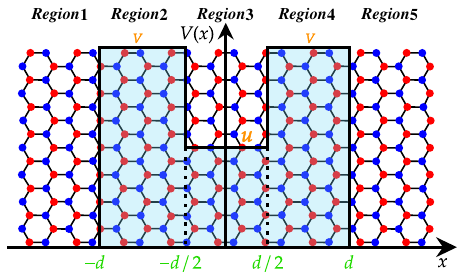}
	\caption{Schematic for the monolayer graphene double barrier.}
	\label{fig:f1}
\end{figure}

We consider a graphene layer model subject to proximity-induced exchange interactions and a double barrier, as shown in Fig. \ref{fig:f1}. The system consists of five regions, denoted as $j = 1, \cdots, 5$, with the potential profile defined over these regions. This setup allows the analysis of quantum transport properties, such as tunneling, reflection, and transmission, under the influence of magnetic and exchange interactions, which have been explored in studies of graphene-based systems with external barriers  \cite{Zheng2022, Fattasse2022}. The proximity effects and barrier configurations significantly affect the electron dynamics, including spin-polarized transmission and group delay time behavior.
\begin{equation} 
	V_j(x)=\left \{
	\begin{array}{ll}
		v, & -d <  x  < -d/2 \\
		u,& -d/2 <  x <  d/2 \\ 
		v, & d/2 <  x  < d \\
		0, & \text{otherwise} \end{array}
	\right.  
\end{equation}

 The system is described by the Hamiltonian \eqref{eq1} to study the low-energy electronic states near the Dirac points 
\begin{equation}\label{eq1}
{H}=
\begin{pmatrix}
\beta\lambda^A_{ex}+\Delta+V_j&   v_F (\tau p_x+i p_y)\\  v_F (\tau p_x-i p_y)&-\beta\lambda^B_{ex}-\Delta+V_j
\end{pmatrix}
\end{equation}
where $v_F$ is the Fermi velocity, $\beta=1(-1)$ for spin up (spin down), and $\tau=1(-1)$ for the K(K') valley. $\beta\lambda^A_{ex}+\Delta$ and $-(\beta\lambda^B_{ex}+\Delta)$ are the band gap edges, with $\lambda^A_{ex}$ and $\lambda^B_{ex}$ denoting the exchange parameters for the sublattices A and B. Here, $\Delta$ is the energy gap induced by the hexagonal boron nitride substrate. {By using $\det(H-E\mathbb{I}_2)$, with $ \mathbb{I}_2$ being the unit matrix, we can}
show that the dispersion relation associated with $H$ is given by
\begin{equation} \label{eq3}
E=V_j+\lambda^-_{ex}+s_j \sqrt{\hbar^2 v_F^2 k_j^2+(\lambda^+_{ex}+\Delta)^2}
\end{equation} 
where $s_j = \text{sgn}(E-V_j-\lambda^-_{ex})$ depending on the sign of the energy difference $E-(V_j+\lambda^-_{ex})$, the wave vector is
\begin{equation}
k_j=\sqrt{k_{xj}^2+k_y^2} 
\end{equation}
 and $\lambda^\pm_{ex}= \beta(\lambda^A_{ex}\pm\lambda^B_{ex})/2$. 
The two components of the wave vector outside the barriers can be written as 
\begin{align}
	k_{xj}= k_j \cos{\phi_j}, \quad k_y= k_j \sin{\phi_j}
\end{align}
and $\phi_1$ is the incident angle.
 Note that the condition 
 \begin{widetext} 
 	\begin{equation} \label{eq4} (E - V_j - \lambda^-{ex})^2 - (\lambda^+{ex} + \Delta)^2 - \left[(E - \lambda^-{ex})^2 - (\lambda^+{ex} + \Delta)^2\right]\sin^2\phi_j < 0 
 \end{equation} 
\end{widetext} 
leads to an exponential decay of the imaginary wave vector in the $y$ direction, indicating that the wave is evanescent inside the barrier. This condition is typical of quantum tunneling and wave propagation through barriers, where the wave does not propagate freely but decays due to the imaginary component of the wave vector.

The symmetry along the translational wave vector $k_y$ allows separating the variables in the eigenspinor associated with the Hamiltonian $H$. This means that the wave function can be written as $\Phi_j(x,y)= \psi_j(x) e^{ik_yy} $. Then, by solving the eigenvalue equation $H \Phi_j(x,y)= E \Phi_j(x,y)$, we obtain the eigenspinors for region 1 ($ x < -d$):
\begin{equation}  
\psi_1=\binom
{1} {z_1}
e^{ik_{x1}x} +r
\binom
{1} {-z_1^*} e^ {-ik_{x1}x}
\end{equation}
region 2 ( $-d <  x < -d/2 $):
\begin{equation} 
\psi_2=\alpha_1
\binom
{1} {z_2}
 e^{ik_{x2}x} +\alpha_2
\binom
{1}
{-z_2^*}
e^ {-ik_{x2}x}
\end{equation}
region 3 ($-d/2 <  x < d/2 $):
\begin{equation}  
\psi_3=\alpha_3\binom
{1} {z_3}
 e^{ik_{x3}x} +\alpha_4\binom
{1} {-z_3^*}
e^ {-ik_{x3}x}
\end{equation}
 region 4  ($d/2 <  x < d $):
\begin{equation}  
\psi_4=\alpha_5\binom
{1} {z_2}
 e^{ik_{x2}x+} +\alpha_6\binom
{1} {-z_2^*}
e^ {-ik_{x2}x}
\end{equation}
and region 5 ( $x > d$):
\begin{equation}  
\psi_5= t \binom
{1} {z_1}
 e^{ik_{x1}x} 
\end{equation}
where the complex number corresponding to region $j$ has the form 
\begin{equation}  
z_j=\frac{\hbar v_F ( \tau k_{xj} -i k_y)}{\beta\lambda^B_{ex}+\Delta+E-V_j} .
\end{equation}
These results will be used to address various aspects of the present system, particularly tunneling effects and group delay time.

\section{Tunneling effect}\label{III}

By applying boundary conditions at the interfaces between the five regions that make up our system, we can explicitly derive the transmission and reflection probabilities. These probabilities allow us to analyze the tunneling effect quantitatively, shedding light on how electrons pass through or reflect from the potential barriers under different conditions. Using this method, we can thoroughly investigate the behavior of the system, such as how the proximity-induced exchange interaction and external potential barriers affect the quantum transport. The numerical results provide important insights into phenomena such as Klein tunneling and spin-dependent group delay time, highlighting the main results of our study. This approach not only improves our understanding of the tunneling dynamics but also allows a more comprehensive discussion of the influence of barrier height, width, and incident electron properties on the transmission and reflection behavior. These results are essential for advancing the application of graphene in spintronic and quantum tunneling devices.

According to the Appendix \ref{BB}, we can arrange the boundary condition results as follows
\begin{equation}  
\binom{1}
{ r}
= M
\binom
{t}  {0}
\end{equation}
with the transfer matrix
\begin{equation}  
M= 
\begin{pmatrix}
M_{11} & M_{12}\\ M_{21} & M_{22}
\end{pmatrix}.
\end{equation}
The transmission  and reflection coefficients are
\begin{align}  
t=\frac{1}{M_{11}}, \quad
r=\frac{M_{21}}{M_{11}}.
\end{align}
The transmission $T=\left|\frac{J_t}{J_i}\right|$ and reflection $R=\left|\frac{J_r}{J_i}\right|$ probabilities are calculated using the current of densities $J_i$, $J_r$, and $J_t$ representing the incident, reflected, and transmitted waves, respectively. We get the current density from the Hamiltonian 
\begin{align}
	J=e\upsilon_F \psi^+\sigma_x\psi
\end{align}
leading to the probabilities
\begin{align}
	T=\left|t\right|^{2}, \quad 	R=\left|r\right|^{2}.
\end{align}	
We shall proceed with numerical analysis after obtaining closed-form equations of the group delay time in various energy domains.

\section{Group delay time}\label{IV}

To gain deeper insight into tunneling dynamics, one can examine the phase time or group delay time. These quantities provide valuable information about the time it takes for a wave packet to propagate through a potential barrier. In addition, a spatially localized wave packet can be constructed by summing over a number of stationary states, each corresponding to different energies. This approach provides a more complete description of the tunneling process, highlighting the contributions of different energy states to the overall dynamics of the wave packet. As a result, we can write \cite{Nimtz, Fattasse2022}
\begin{equation}  
\psi(x,t)=\int_{E} f(E-E_0) \psi(E) e^{-i Et/\hbar}dE
\end{equation}
and $f(E-E_0)$ is an energy distribution such as a Gaussian centered at the mean energy $E_0$ \cite{Beenakker}. The wave packet divided into two reflected and transmitted waves
\begin{align} 
&
\psi_r(x,t)=\int_{E} f(E-E_0)
r(E) \binom
{1}  {-z_1^*}
 e^{i(-k_{x1}x+k_yy)}
e^{-i \frac{Et}{\hbar}}dE\\ 
&
\psi_t(x,t)=\int_{E} f(E-E_0)
t(E) \binom
{1} {z_1}
 e^{i(k_{x1}x+k_yy)}
e^{-i \frac{Et}{\hbar}}dE
\end{align}
where  the amplitudes of reflection and  transmission can be written as 
\begin{align}\label{20}
&t(E)= 
\frac{1}{\left|M_{11}\right|} e^{i \phi_t(E)}\\	
	&r(E) = 
\left|\frac{M_{21}}{M_{11}}\right| e^{i \phi_r(E)}\label{21}.
\end{align} 

By applying the stationary phase approximation, we can obtain analytical expressions for the group delay time, assuming that the distribution \( f(E-E_0) \) is smooth and sharply peaked around the central energy or wavevector. This approximation simplifies the evaluation of the integrals by focusing on the region where the phase of the integrand changes most slowly, corresponding to the peak of \( f(k_y, \omega) \) \cite{Steinberg1, Li1}.  Based on this  method, we find the  phase time of the transmitted wave  at ($t=0,x=2 d$)
\begin{equation}  
\tau_{gt}=\hbar \frac{\partial}{\partial E}(\phi_t+2 k_{x1} d)
\end{equation}
and similarly  for  the reflected wave   at ($t=0,x=-2 d$) 
\begin{equation}  
\tau_{gr}=\hbar \frac{\partial}{\partial E}(\phi_r+2 k_{x1} d).
\end{equation}
The phase time takes the form
\begin{equation}  
\tau_g=|t|^2\tau_{gt}+|r|^2\tau_{gr}
\label{eq26}.
\end{equation}
In the Appendix \ref{AA}, we explicitly show that the transmission \eqref{20} and reflection \eqref{21} coefficients can be expressed in terms of different parameters as 
\begin{align}  
&	t=\frac{A}{\sqrt{a_1^2+ b_1^2}}   e^{i \phi_t(E)}  \\
&  r=\frac{(a_2a_3+ b_2b_3)^2 +(a_3b_2 -a_3b_2)^2 }{(a_3^2+ b_3^2)^2}e^{i \phi_r(E)}
	\label{eq27}
\end{align}
and the corresponding phases take the forms
\begin{align}   
&	\phi_t=\arctan\left(-\frac{b_1}{a_1}\right) \\
	& \phi_r=\arctan\left(\frac{a_3 b_2-a_2 b_3}{a_2 a_3+b_2 b_3}\right).
\end{align} 
To emphasize our results and provide a clearer understanding of the behavior of the system, we proceed with numerical simulations to highlight the key features of our model. These calculations allow us to study various quantum transport properties, such as transmission, reflection, and group delay time, under the influence of proximity-induced exchange interactions and double-barrier structures. By simulating different parameter values, including barrier heights, widths, and incident angles, we can better understand how these factors affect tunneling dynamics, especially in the context of graphene-based systems.

\section{Numerical results}\label{V}

\begin{figure*} [ht] 
	\centering 
	\subfloat[]	{\includegraphics[scale=0.45]{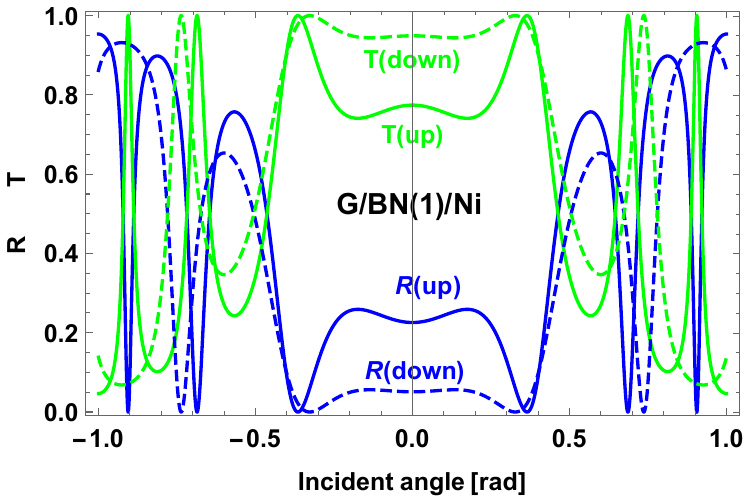}}
	\subfloat[]	{\includegraphics[scale=0.45]{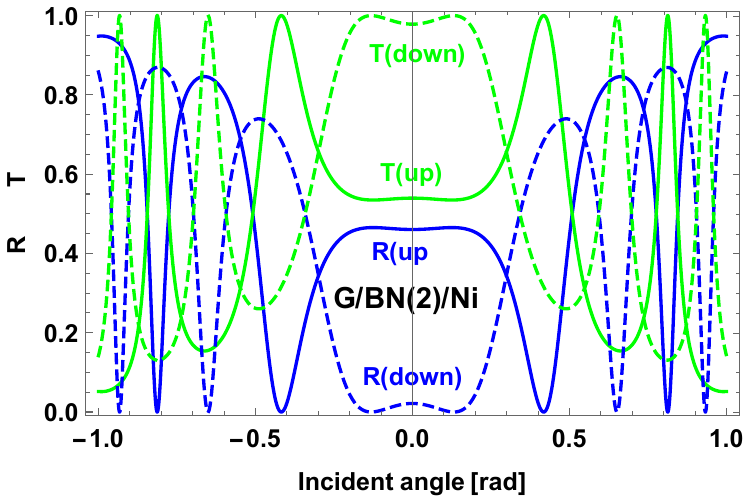}}
	\subfloat[]	{\includegraphics[scale=0.45]{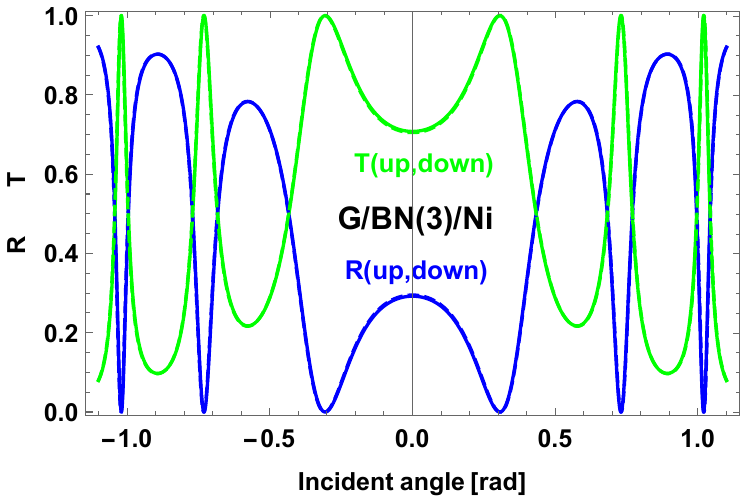}}\\
	\subfloat[]	{\includegraphics[scale=0.45]{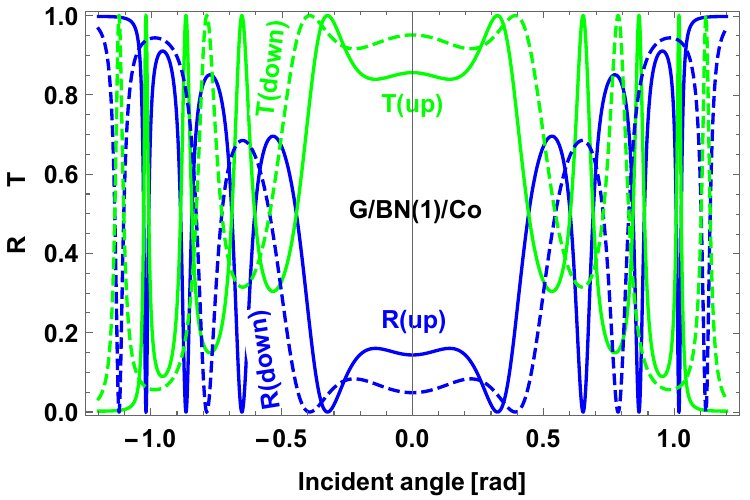}}
	\subfloat[]	{\includegraphics[scale=0.45]{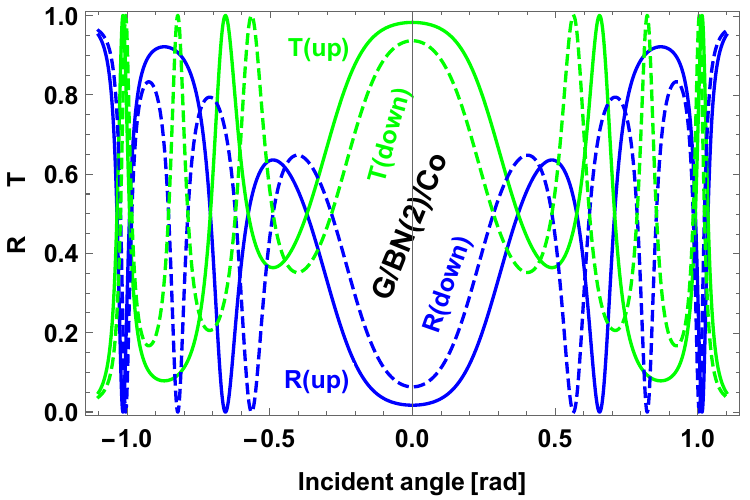}}
	\subfloat[]	{\includegraphics[scale=0.45]{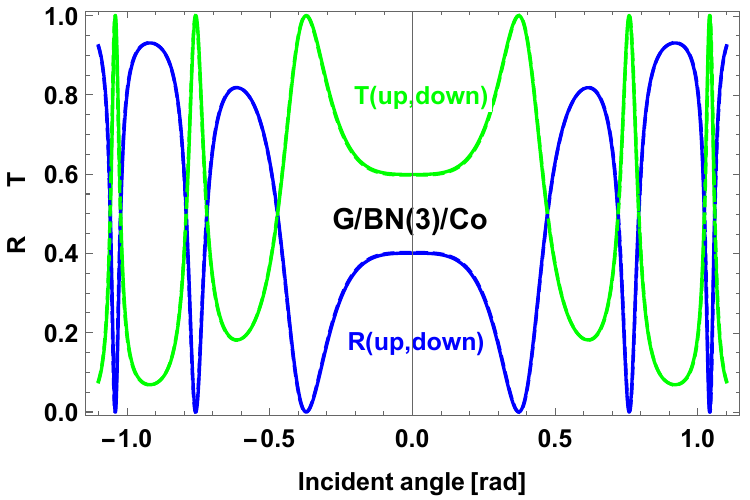}}
	\caption{The transmission $T$ (green) and reflection $R$ (blue) probabilities as a function of the incident angle $\phi_1$ for spin-up and spin-down (dashed lines) with 
		{(a,b,c,d,e,f) correspond to  the parameter configurations, respectively}: $\Delta=22.86,42.04,40.57,19.25,36.44,38.96$ meV, $\lambda_{ex}^A=-1.4,0.068,-0.005,-3.14,0.097,-0. 005$ meV, $\lambda_{ex}^B~=~7.78,-3.38,0.017,8.59,-9.81,0.018$ meV, $v_F=810,824,826,812,820,821$ km/s. Here we choose the barrier heights $v=400$ meV and $u=200$ meV, the incident energy  $E=100 $ meV and the barrier width $d= 80$ nm.}
	\label{fig:f2} 
\end{figure*} 

To gain a clearer understanding of the transmission and phase-time behavior in the presence of potential barriers, we numerically compute the transmission and reflection probabilities, as well as the group delay time, under various conditions. These conditions include incident angle, incident energy, and barrier length, considering different system parameters for a graphene monolayer/insulator (hBN)/ferromagnetic metallic substrate (Ni or Co). We study both spin-up and spin-down orientations to investigate how the spin configuration affects these properties. This approach allows us to capture the effect of different system configurations on quantum transport in graphene, similar to the studies \cite{Zheng2022,Fattasse2022}. The results help to elucidate how barrier properties affect tunneling dynamics  and shed light on the role of proximity effects in spintronic applications of 2D materials. {For the numerical analysis, 
we use the table \ref{tab1}, which gives the parameter values for the graphene (G), boron nitride (BN), and cobalt/nickel (Co/Ni) systems, including the exchange parameters ($\lambda_{ex}^A$, $\lambda_{ex}^B$), the band gap $\Delta$, and the Fermi velocity $v_F$. These values were obtained by Zollner {\it et al.} \cite{Zollner} via ab-initio calculations.
}

\begin{table}[h]
\centering
	\begin{tabular}{|l|c|c|c|c|}
		\hline
		System & $\lambda_{ex}^A$ [meV] & $\lambda_{ex}^B$ [meV] & $\Delta$ [meV] & $v_F$ [km/s] \\
		\hline
		$\mathrm{G} / \mathrm{BN}(1) / \mathrm{Ni}$ & -1.40 & 7.78 & 22.86 & 810 \\
		$\mathrm{G} / \mathrm{BN}(2) / \mathrm{Ni}$ & 0.068 & -3.38 & 42.04 & 824 \\
		$\mathrm{G} / \mathrm{BN}(3) / \mathrm{Ni}$ & -0.005 & 0.017 & 40.57 & 826 \\
		\hline
		$\mathrm{G} / \mathrm{BN}(1) / \mathrm{Co}$ & -3.14 & 8.59 & 19.25 & 812 \\
		$\mathrm{G} / \mathrm{BN}(2) / \mathrm{Co}$ & 0.097 & -9.81 & 36.44 & 820 \\
		$\mathrm{G} / \mathrm{BN}(3) / \mathrm{Co}$ & -0.005 & 0.018 & 38.96 & 821 \\
		\hline
	\end{tabular}
	\caption{Parameters of the  Hamiltonian \eqref{eq1} for $\mathrm{G} / \mathrm{BN}(\mathrm{i}) / \mathrm{Ni}$ and $\mathrm{G} / \mathrm{BN}(\mathrm{i}) / \mathrm{Co}$ systems, with $i=1,2,3$ denotes layers.}\label{tab1}
\end{table}

Fig. \ref{fig:f2} shows the transmission and reflection probabilities as a function of incident angle for both spin-up and spin-down orientations (dashed lines). The potential barrier heights are set to $v=400$ meV and $u=200$ meV, while the energy of the incident electrons is $100$ meV. The primary goal is to observe how the proximity exchange interaction and the number of BN atomic levels affect these probabilities in a double barrier system with different ferromagnetic metallic substrates, specifically nickel (Ni) and cobalt (Co). From the results in Fig. \ref{fig:f2}, several important observations can be made. First, there are peaks corresponding to ideal transmission at certain {incident angles}, indicative of Klein tunneling. In particular, near normal incidence, the transmission probability is less than one. In addition, the transmission behavior is spin dependent, with a clear shift between spin-up and spin-down states, {see Figs. \ref{fig:f2}a,b,c,d}. In the case of the G/BN(3)/Ni and G/BN(3)/Co structures {(Figs. \ref{fig:f2}e,f)}, however, the proximity exchange coupling vanishes, leading to spin-independent transmission and reflection probabilities. The symmetry at normal incidence is also evident, and the total probability satisfies $R+T=1$, confirming the conservation of probability. Finally, compared to the single-barrier case, the double-barrier structure exhibits an increased number of transmission oscillation peaks, in agreement with previous findings \cite{Tepper}.

\begin{figure*}  [ht]  \centering
\subfloat[]	{\includegraphics[scale=0.45]{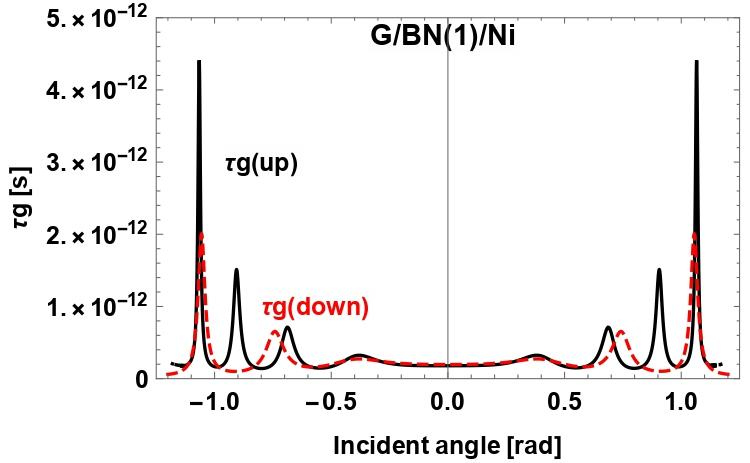}} 
\subfloat[]	{\includegraphics[scale=0.45]{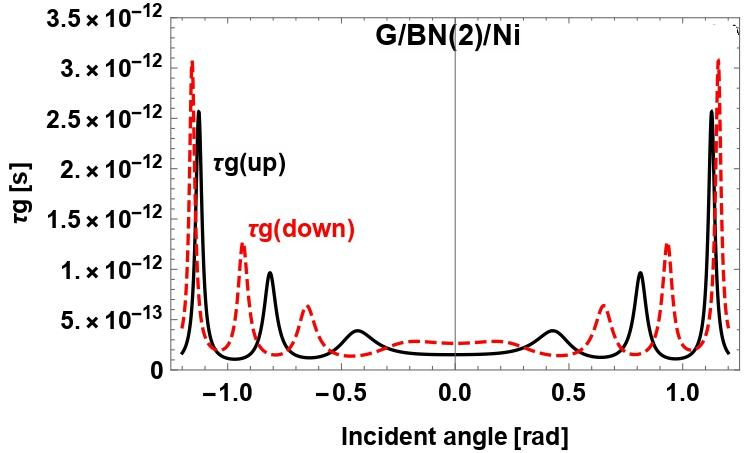}}
\subfloat[]	{\includegraphics[scale=0.45]{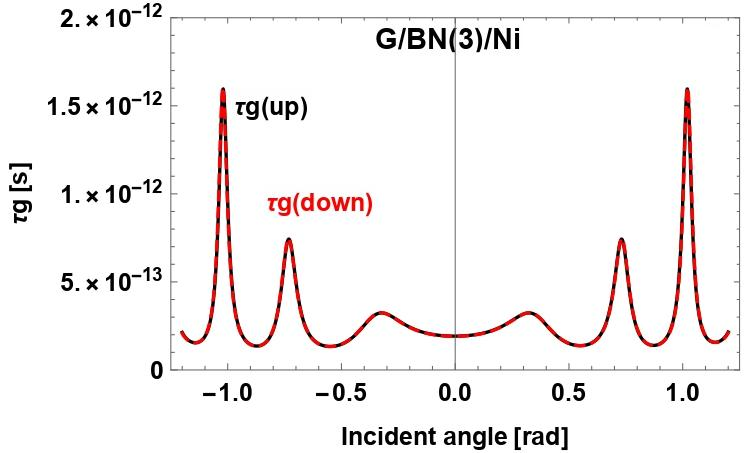}}\\
\subfloat[]	{\includegraphics[scale=0.45]{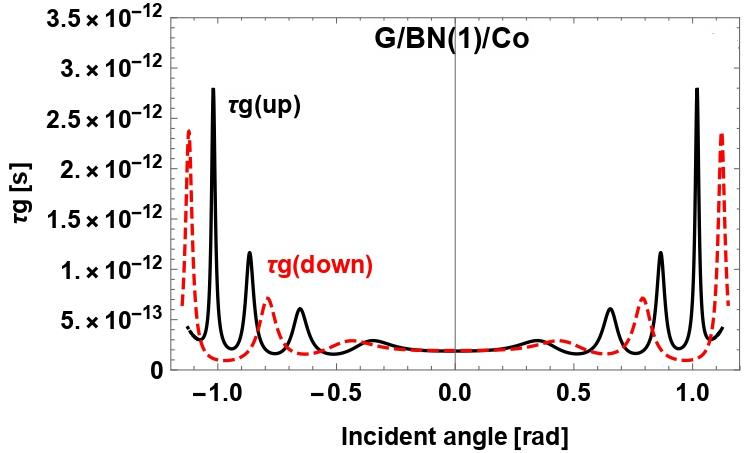}}
\subfloat[]	{\includegraphics[scale=0.45]{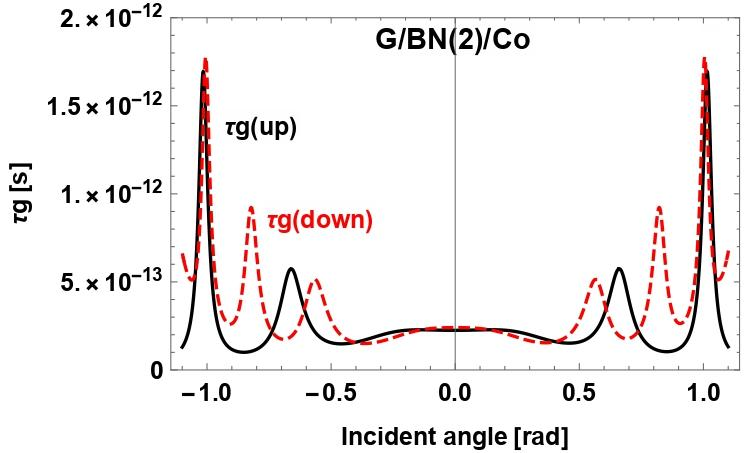}}
\subfloat[]	{\includegraphics[scale=0.45]{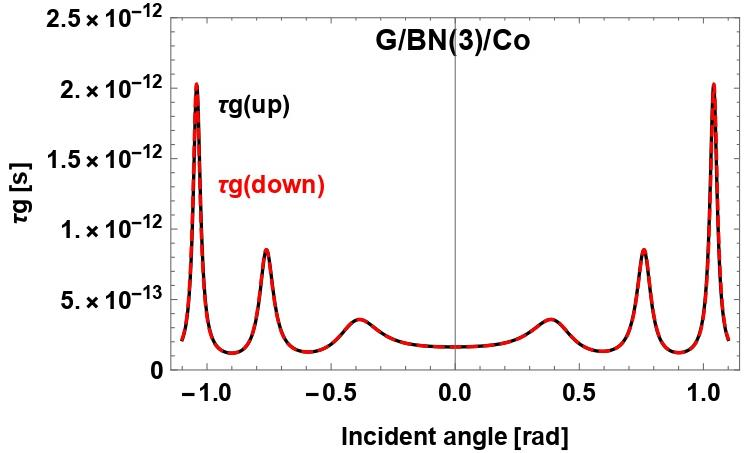}}
	\caption{The group delay time $\tau_g$ as a function of incident angle $\phi_1$ for spin-up and spin-down (red dashed lines) with the same configurations as in Fig. \ref{fig:f2}.}
	\label{fig:f3} 
\end{figure*}

It is interesting to analyze how the group delay time, as modeled by equation \eqref{eq26}, varies with the incident angle under proximity exchange coupling. The numerical results are shown in Fig. \ref{fig:f3}, where the potential barrier heights are set to $v=400$ meV and $u=200$ meV, and the incident electron energy is 100 meV. First, we observe that the group delay time exhibits peaks at the same incident angles where transmission resonances occur in Fig. \ref{fig:f2}. This suggests that the number of phase time peaks corresponds directly to the transmission resonances, with the group delay time maxima coinciding with the tunneling probability peaks. In addition, the group delay time shows a spin-dependent nature, changing its behavior depending on the spin orientation, {see~Figs. \ref{fig:f2}a,b,d,e}. However, {in Figs. \ref{fig:f3}c,f,} the {spin-up} and spin-down behaviors become identical, indicating the absence of proximity exchange coupling after three BN monolayers. In addition, the group delay time is larger in the G/BN(1,2)/Co {(Figs. \ref{fig:f3}d,e)} structure than in the G/BN(1,2)/Ni case {(Figs. \ref{fig:f3}a,b)}, but this trend is reversed for the G/BN(3)/Co {(Fig. \ref{fig:f3}f)} structure, where the delay becomes smaller than in the G/BN(3)/Ni case {(Figs. \ref{fig:f3}c)}. {In general}, the group delay time behaves similarly to the transmission probability, with symmetry at normal incidence. Compared to the single-barrier scenario, the group delay time is generally larger, with more pronounced peaks over the same range of incident angles, consistent with previous observations \cite{Tepper}.

\begin{figure*} [ht]
	\centering
\subfloat[]{	\includegraphics[scale=0.36]{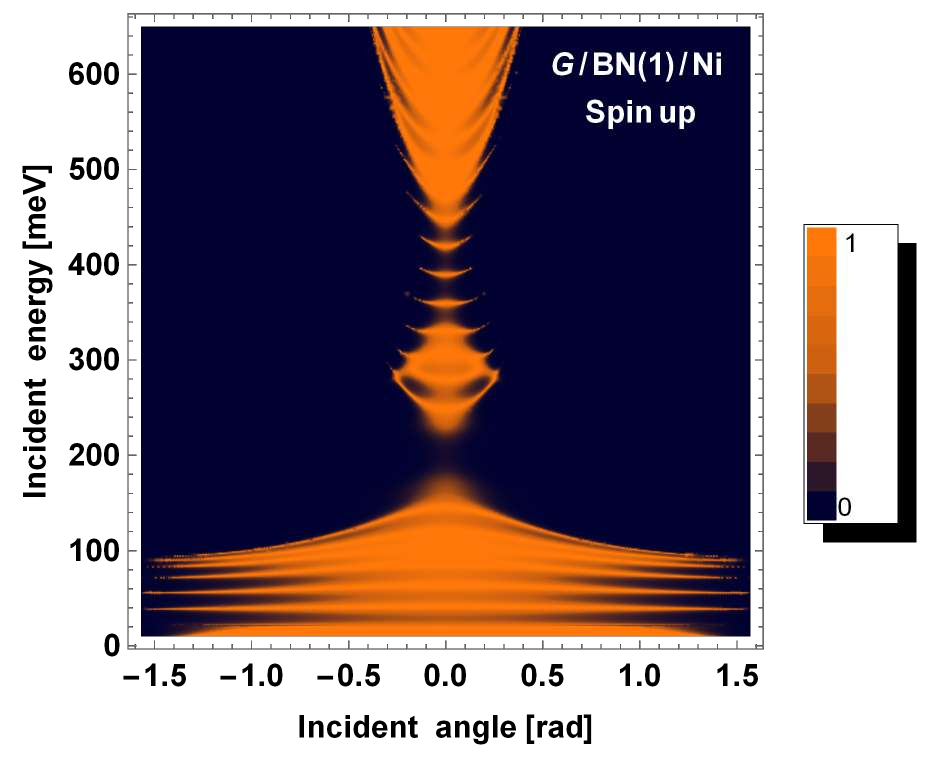}}
\subfloat[]	{\includegraphics[scale=0.36]{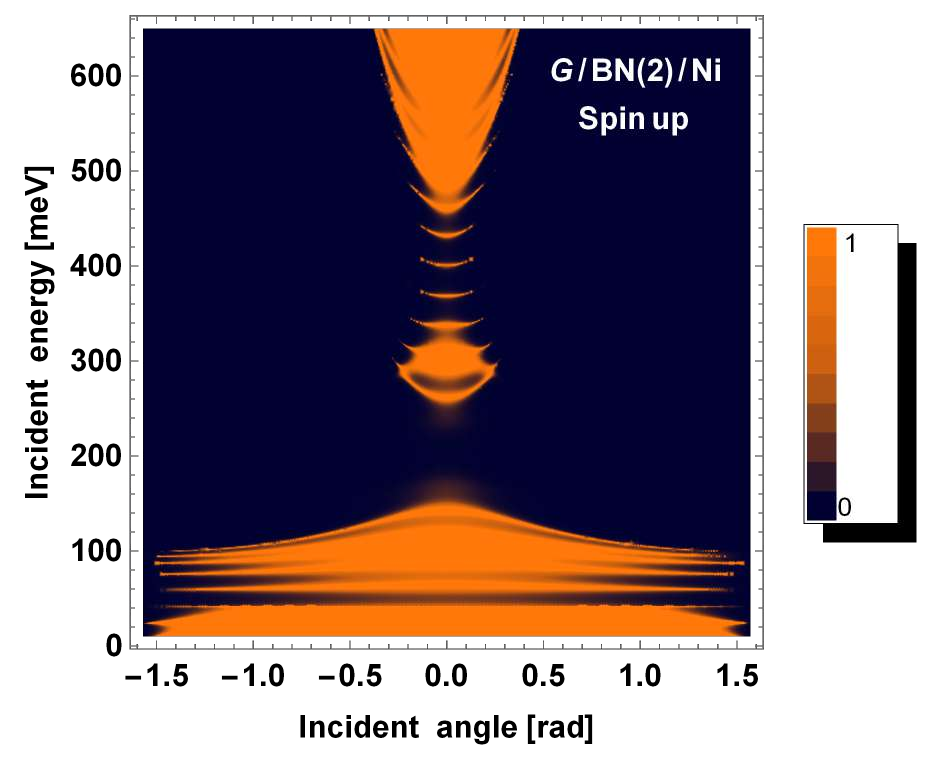}}
\subfloat[]	{\includegraphics[scale=0.36]{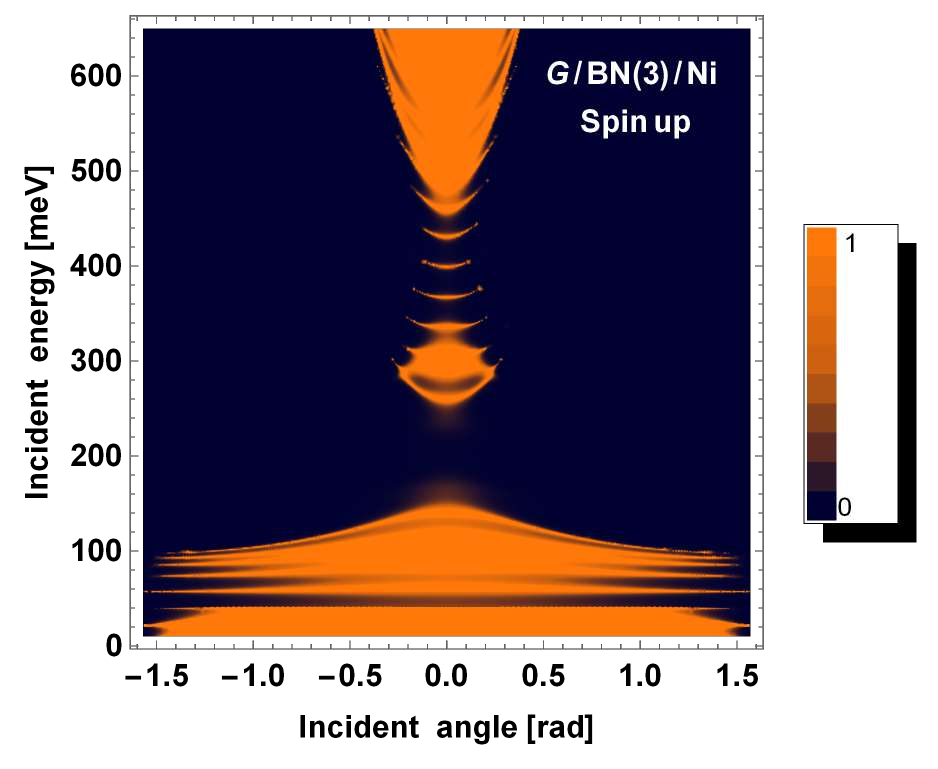}}\\
\subfloat[]	{\includegraphics[scale=0.36]{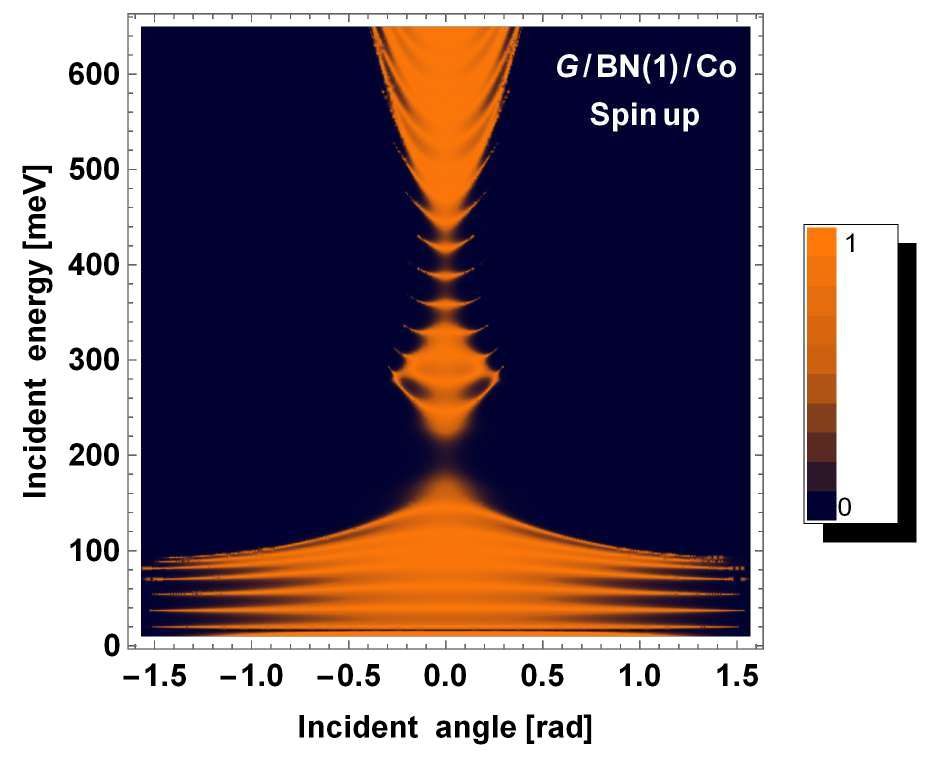}}
\subfloat[]	{\includegraphics[scale=0.36]{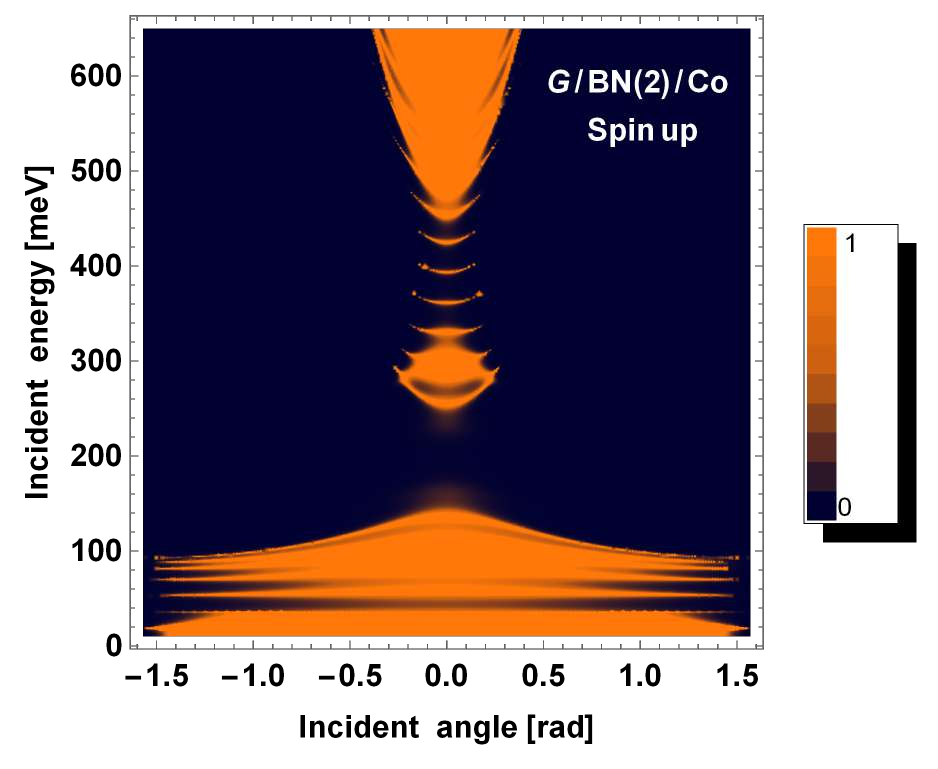}}
\subfloat[]	{\includegraphics[scale=0.36]{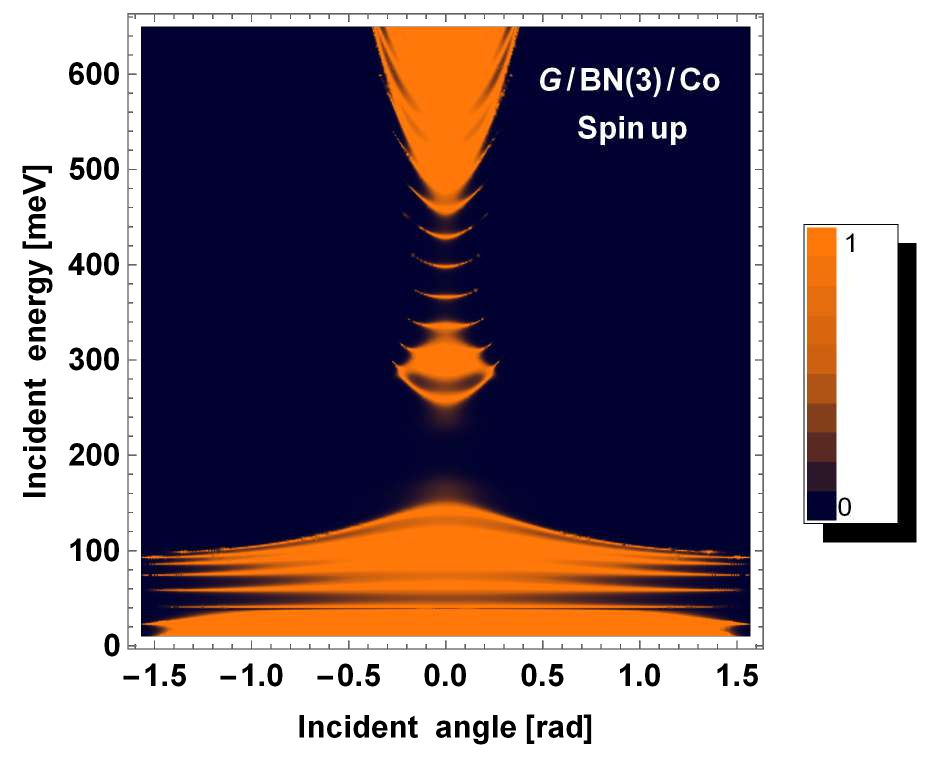}}
	\caption{ Density plot of transmission probability $T$ as a function of the incident angle $\phi_1$ and incident energy $E$ for spin-up orientation 
	with the same configurations as in Fig. \ref{fig:f2} except the barrier width is $d= 40$ nm.}
	\label{fig:f4}
\end{figure*}

Fig. \ref{fig:f4} shows a density plot of the transmission probability as a function of both incident angle and energy for spin-up electrons with potential barrier heights of $v=400$ meV, $u=200$ meV, and a barrier width of $d=40$ nm. We observe two Dirac points at $E = u + \lambda_{ex}^-$ and $E = v + \lambda_{ex}^-$, which represent different energy thresholds for the transmission behavior. In the lower energy regime ($E < u + \lambda_{ex}^-$), full transmission is observed at certain energies, even when the particle energy is below the barrier height. At the first Dirac point ($E = u + \lambda_{ex}^-$), the transmission drops to zero, and there are no resonances. In the intermediate energy range ($u + \lambda_{ex}^- < E < v + \lambda_{ex}^-$), we observe distinct transmission peaks, indicating resonant tunneling through the barriers. At the second Dirac point ($E = v + \lambda_{ex}^-$), the transmission is mostly suppressed, although resonance peaks appear due to the bound states of the double barrier. For energies above the second Dirac point ($E > v + \lambda_{ex}^-$), Dirac fermions show pronounced transmission resonances. Comparing the transmission in G/BN(i)/Co structures {(Figs. \ref{fig:f4}d,e,f)} with G/BN(i)/Ni {(Figs. \ref{fig:f4}a,b,c)}, we see sharper peaks around the second Dirac point for G/BN(i)/Co, {$i=1,2,3$ indicates layers. In our system, a graphene layer is deposited on a few atomic layers of boron nitride (BN), a wide-gap insulator. The underlying ferromagnetic metal substrate (Co or Ni) induces a uniform exchange field in graphene \cite{Ba2017}. However, the proximity exchange effect—comparable to the interlayer exchange coupling in magnetic multilayers—is significantly reduced when three BN layers are used \cite{Tepper}. This reduction of the spin-dependent influence explains why the group delay time becomes spin-independent in Figs. \ref{fig:f4}c,f.} Moreover, compared to the single barrier case, the double barrier setup introduces two Dirac points and transmission peaks in the bandgap that are absent in the single barrier configuration \cite{Tepper}.

\begin{figure*}[ht]
	\centering
\subfloat[]	{\includegraphics[scale=0.43]{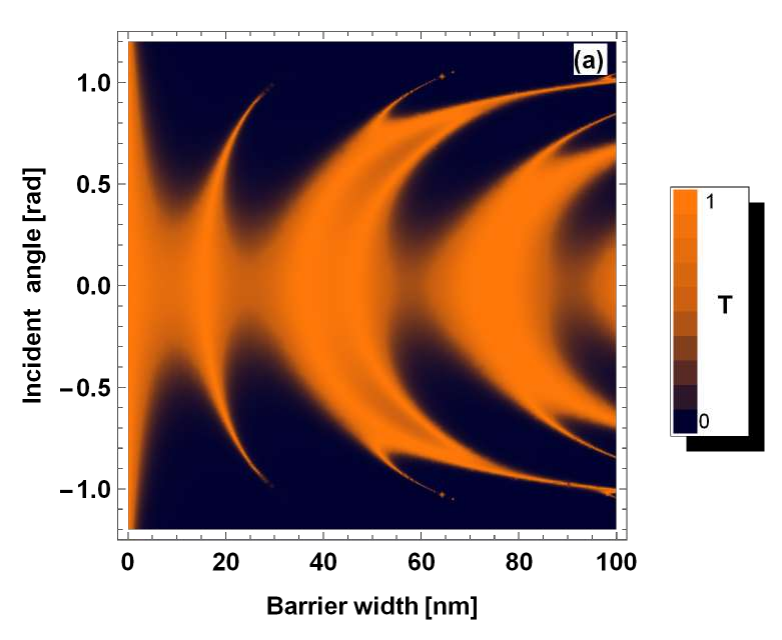}}
\subfloat[]	{\includegraphics[scale=0.43]{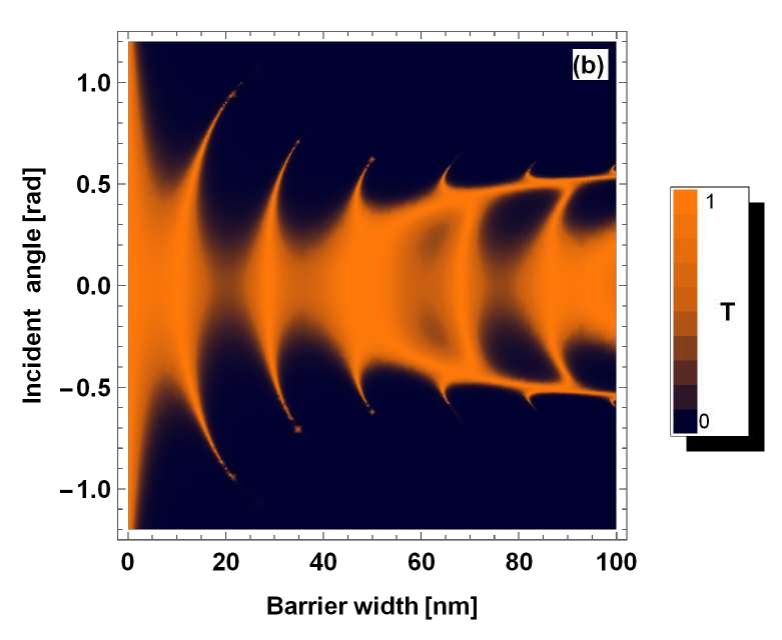}}
\subfloat[]	{\includegraphics[scale=0.43]{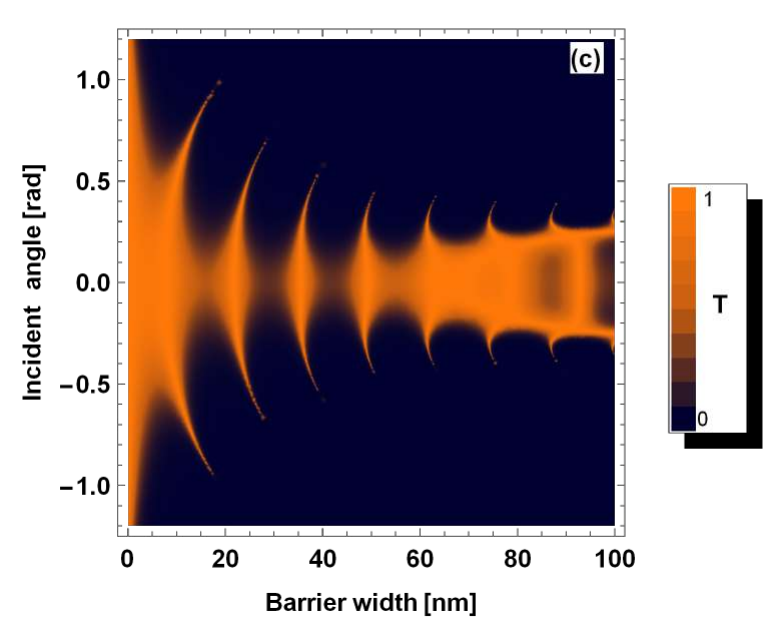}}\\
\subfloat[]	{\includegraphics[scale=0.43]{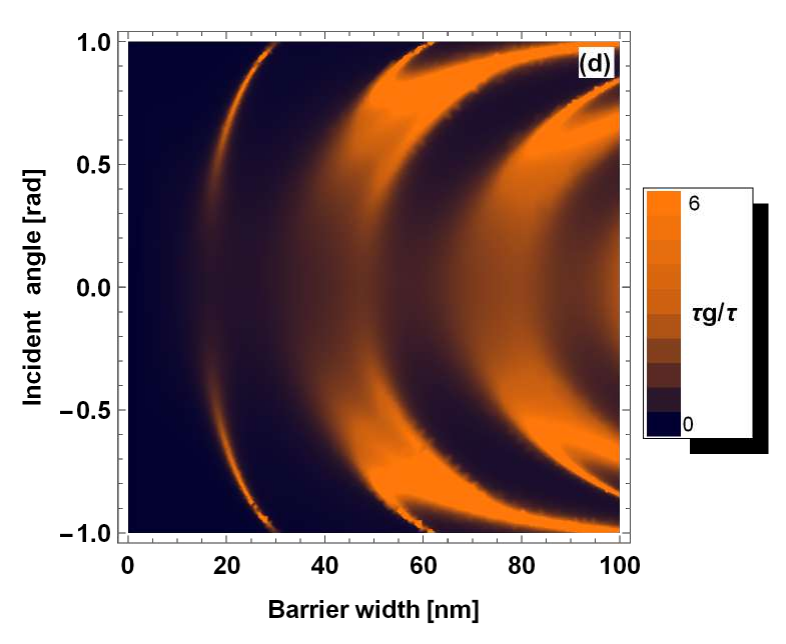}}
\subfloat[]{	\includegraphics[scale=0.43]{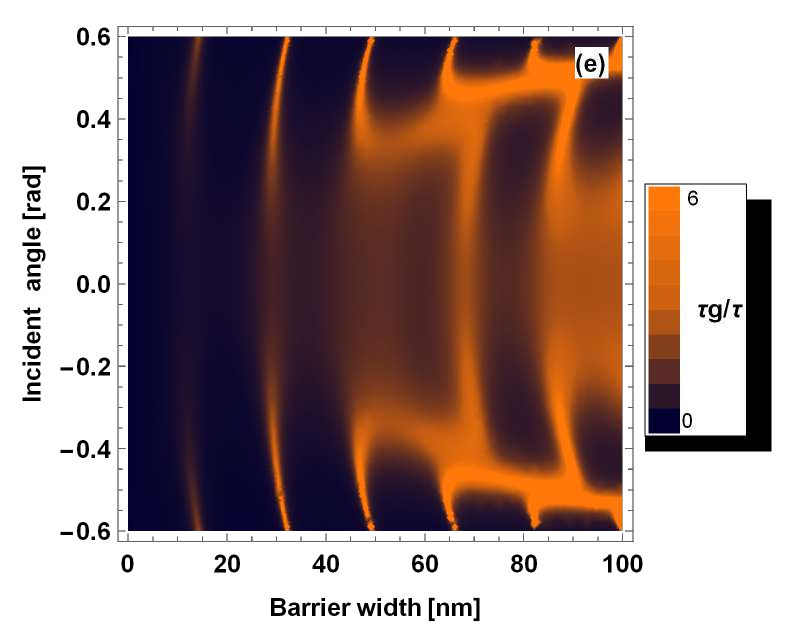}}
\subfloat[]	{\includegraphics[scale=0.43]{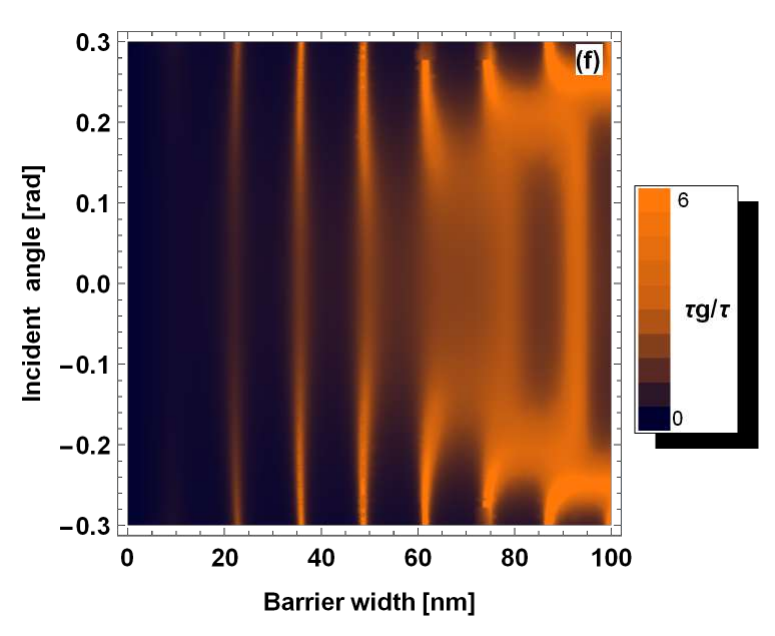}}
	\caption{Density plot of transmission probability $T$ and phase time $\tau_g$ as a function of incident angle $\phi_1$ and barrier width $d$ for electron spin-up in the G/BN(1)/Ni system ($\Delta=22. 86$ meV, $\lambda_{ex}^A= -1.4$ meV ,$\lambda_{ex}^B= 7.78$ meV, $v_F=810$ km/s), the barrier heights are $v=200$ meV and $u=0$ meV. The first line describes $T$ and the second line is for $\tau_g$. Here the incident energies $E= 100, 120, 140$ meV correspond to the first, second and third columns, respectively. 
	}
	\label{fig:f5}
\end{figure*}

Let us now explore how barrier thickness affects transmission probability group delay time, as well as the importance of evanescent conditions. Fig. \ref{fig:f5} shows the transmission probability and group delay time as a function of barrier height and incident angle. The potential barrier heights are set to $v=200$ meV and $u=0$ meV, with incident energies of $E=100$ meV, $E=120$ meV, and $E=140$ meV. By examining the evanescence condition derived from equation \eqref{eq4}, we can identify the critical incident angles at which the wave vector inside the barrier becomes imaginary. For $E=100$ meV the critical angle is $\phi_{cr1}=1.13403$ rad, for $E=120$ meV it is $\phi_{cr2}=0.619764$ rad, and for $E=140$ meV the critical angle is $\phi_{cr3}=0.351021$ rad. As the energy increases, the number of transmission peaks also increases {(Figs. \ref{fig:f5}a,b,c)}, satisfying the evanescence condition, with the maximum energy required to observe an imaginary wave vector inside the barrier being $E=169.35$ meV. In addition, as shown in {Figs. \ref{fig:f5}d,e,f}, the group delay time oscillates with increasing barrier width and shows no signs of saturation—indicating the absence of the Hartman effect. This contrasts with the results of the single-barrier case, where the group delay time was found to saturate with increasing barrier width \cite{Tepper}. In our case, however, the group delay time grows with increasing barrier width, highlighting the distinct dynamics present in the double-barrier structure.

\section{Conclusion}\label{VI}
 
{Our work built on the results developed in \cite{Tepper} and extended them to the case of a double rectangular barrier. First, we solved the Dirac equation and obtained the solutions of the energy spectrum over five different regions. Using the solutions at the boundaries, we derived the transmission and reflection coefficients using the transfer matrix method. In addition, we studied the group delay time by analyzing a Gaussian wave packet centered at a specific point. Then, we applied the stationary phase method to extract the transmission and reflection phase times. The numerical analysis was then discussed in detail, focusing on Klein tunneling and group delay time, studying their dependence on factors such as incident angle, barrier height, barrier width, and the energy of the incident electrons. Spin-up and spin-down orientations were also considered to highlight the spin-dependent dynamics in the system.}

We have found that in the case of a double barrier, both transmission and group delay time exhibit spin-dependent behavior, with noticeable shifts between spin-up and spin-down orientations, especially up to three levels of boron nitride (BN). The number of Klein tunneling peaks increases in the double-barrier structure, and full transmission is observed for certain values of incident angle, barrier width, and electron energy. In addition, the group delay time becomes significantly longer in the double-barrier structure compared to the single-barrier case. Our analysis also highlights the importance of the evanescent state, where we identify the critical angle of incidence and the maximum energy for the system depending on the values of the potential barriers. These results provide a deeper insight into the interaction of graphene with proximity exchange fields in a double-barrier configuration.

Finally, by setting the values \(v = u = V\) and \(d = D/2\) in our generalized formulation, we recover the results from the single-barrier model presented in \cite{Tepper}. This shows that our approach not only reproduces known results but also extends them to more complex double-barrier structures, confirming the broader applicability and robustness of our formulation. Thus, our results provide a more comprehensive and generalized understanding of the behavior of the system, especially in terms of transmission, group delay time, and Klein tunneling under proximity exchange effects.

\section*{Acknowledgment}
P.D. and D.L. acknowledge partial financial support from FONDECYT 1231020.

\vspace{1cm}
\appendix\label{Appendix}
\section{Determining transmission and reflection}\label{BB}
To obtain the transmission and reflection probabilities, we use the boundary conditions of the eigenspinors at the interfaces $x=-d, -d/2, d/2, d$. These are
\begin{align}
&\psi_1(-d)	=\psi_2(-d)\\
&\psi_2(-d/2)	=\psi_3(-d/2)\\
&\psi_3(d/2)	=\psi_4(d/2)\\
&\psi_4(d)	=\psi_5(d).
\end{align}
{To go further, it is convenient to use the matrix representation. As a result, we can express the set of equations as follows}
\begin{align}
&	M_1(-d)\binom{1}{r}
	= M_2(-d)
	\binom{	\alpha_1 }{ \alpha_2}
\\
&	M_2(-d/2)
	\binom
		{\alpha_1}  {\alpha_2}
	= M_3(-d/2)	
	\binom
		{\alpha_3}  {\alpha_4}
\\
&
	M_3(d/2)
	\binom{\alpha_3}  {\alpha_4}
= M_2(d/2)
		\binom
		{\alpha_5}  {\alpha_6}
\\
&
	M_2(d)\begin{pmatrix}
		\alpha_5 \\ \alpha_6
	\end{pmatrix}= M_4(d)
	\binom
	{t} {0}
\end{align}
and  the matrices have the forms 
\begin{align}
	& M_1(-d)=\begin{pmatrix}
		e^{-i k_{x1} d} & e^{i k_{x1} d} \\ z_1  e^{-i k_{x1} d} & -z_1^* e^{i k_{x1} d}
	\end{pmatrix} 
	\\
	&
	M_2(-d)=\begin{pmatrix}
		e^{-i k_{x2} d} & e^{i k_{x2} d} \\ z_2  e^{-i k_{x2} d} & -z_2^* e^{i k_{x2} d}
	\end{pmatrix} 
	\\ &
	M_2(-d/2)=\begin{pmatrix}
		e^{-i k_{x2} d/2} & e^{i k_{x2} d/2} \\ z_2  e^{-i k_{x2} d/2} & -z_2^* e^{i k_{x2} d/2}
	\end{pmatrix} \\
	&
	M_3(-d/2)=\begin{pmatrix}
		e^{-i k_{x3} d/2} & e^{i k_{x3} d/2} \\ z_3  e^{-i k_{x3} d/2} & -z_3^* e^{i k_{x3} d/2}
	\end{pmatrix}
	\\ &
	M_3(d/2)=\begin{pmatrix}
		e^{i k_{x3} d/2} & e^{-i k_{x3} d/2} \\ z_3  e^{i k_{x3} d/2} & -z_3^* e^{-i k_{x3} d/2}
	\end{pmatrix}\\
	&
	M_2(d/2)=\begin{pmatrix}
		e^{i k_{x2} d/2} & e^{-i k_{x2} d/2} \\ z_2  e^{i k_{x2} d/2} & -z_2^* e^{-i k_{x2} d/2}
	\end{pmatrix}
	\\ &
	M_2(d)=\begin{pmatrix}
		e^{i k_{x2} d} & e^{-i k_{x2} d} \\ z_2  e^{i k_{x2} d} & -z_2^* e^{-i k_{x2} d}
	\end{pmatrix}
	\\
	&
	M_4(d)=\begin{pmatrix}
		e^{i k_{x1} d} & 0 \\ z_1  e^{i k_{x1} d} & 0
	\end{pmatrix} .
\end{align}

{To determine the transmission and reflection coefficients, we can combine all of the above matrices to form an equation involving only input and output. This is}
\begin{equation}  
	\binom{1}
	{ r}
	= M
	\binom
	{t}  {0}
\end{equation}
{where  the transfer matrix is given by }
\begin{widetext}
	\begin{equation}  
		M= M_1^{-1}(-d) M_2(-d) M_2^{-1}(-d/2) M_3(-d/2) M_3^{-1}(d/2) M_2(d/2) M_2^{-1}(d) M_4(d)
		= 
		\begin{pmatrix}
			M_{11} & M_{12}\\ M_{21} & M_{22}
		\end{pmatrix}.
	\end{equation}
\end{widetext}
The transmission  and reflection coefficients are
\begin{align}  
	t=\frac{1}{M_{11}}\\
	r=\frac{M_{21}}{M_{11}}.
\end{align}
{It is important to note that the matrix elements include all relevant physical parameters $(E, \lambda_{ex}^A$, $\lambda_{ex}^B, \Delta, v_F)$.}

\section{Computing group delay time}\label{AA}

To derive the group delay time, we first write the transmission and reflection coefficients as complex numbers. Then, after some extensive algebra, we show the result
\begin{align}  
&	t=\frac{A}{a_1+i b_1} \\
&	  r=\frac{a_2+i b_2}{a_3+i b_3}
	\label{eq27}
\end{align}
with the quantity \begin{equation}  
	A=16 \,  \frac{k_{x1}}{A_1}\left(\frac{k_{x2}}{A_2}\right)^2  \frac{k_{x3}}{A_3} .
\end{equation}
The phases can be expressed as
\begin{align}   
&	\phi_t=\arctan\left(-\frac{b_1}{a_1}\right) \\
	& \phi_r=\arctan\left(\frac{a_3 b_2-a_2 b_3}{a_2 a_3+b_2 b_3}\right)
\end{align} 
where the different quantities are
\begin{widetext}
	\begin{align}
		a_1 =& -(\chi_1^2+\chi_2^2) \cos[d(5 k_{x1}/2- k_{x2}/2+k_{x3})] -(\chi_3^2+\chi_4^2) \cos[d(5 k_{x1}/2+3 k_{x2}/2+k_{x3})] \nonumber \\
		& +2(\chi_1 \chi_3+\chi_2 \chi_4)\cos[d(5 k_{x1}/2+ k_{x2} /2+k_{x3})]+(\chi_5^2+\chi_6^2) \cos[d( 5 k_{x1}/2- k_{x2}/2-k_{x3})]
		\\
		&
		+(\chi_7^2+\chi_8^2) \cos[d( 5 k_{x1}/2+3 k_{x2}/2-k_{x3})]+ 2 ( \chi_5 \chi_7+  \chi_6 \chi_8) \cos[d(5 k_{x1}/2+k_{x2}/2-k_{x3})]   \nonumber
		\\
		b_1 =& -(\chi_1^2+\chi_2^2) \sin[d(5 k_{x1}/2- k_{x2}/2+k_{x3})] -(\chi_3^2+\chi_4^2) \sin[d(5 k_{x1}/2+3 k_{x2}/2+k_{x3})] \nonumber \\
		& +2(\chi_1 \chi_3+\chi_2 \chi_4)\sin[d(5 k_{x1}/2+ k_{x2} /2+k_{x3})]+(\chi_5^2+\chi_6^2) \sin[d( 5 k_{x1}/2- k_{x2}/2-k_{x3})]
		\\
		&
		+(\chi_7^2+\chi_8^2) \sin[d( 5 k_{x1}/2+3 k_{x2}/2-k_{x3})]+ 2 ( \chi_5 \chi_7+  \chi_6 \chi_8) \sin[d(5 k_{x1}/2+k_{x2}/2-k_{x3})]   \nonumber
		\\  
		a_2=& (\chi_1 \chi_7+  \chi_2 \chi_8) (\cos[2 d(k_{x3}-k_{x1})]-\cos[2 d(k_{x2}-k_{x1})])
		-(\chi_2 \chi_7-\chi_1 \chi_8)(\sin[2d (k_{x3}-k_{x1})]\nonumber \\
		& -\sin[2d (k_{x2}-k_{x1})]) 
		+(\chi_3 \chi_7+ \chi_4 \chi_8 ) (\cos[ d(- 2 k_{x1}+k_{x2})]-\cos[ d(- 2 k_{x1}+k_{x2}+2 k_{x3})])\nonumber\\
		& -(\chi_4 \chi_7- \chi_3 \chi_8 )(\sin[d(- 2 k_{x1}+k_{x2})]-\sin[d (- 2 k_{x1}+k_{x2}+2 k_{x3})])+(\chi_1 \chi_5
		+  \chi_2 \chi_6)  
		\\ & (\cos[d(- 2 k_{x1}+k_{x2}+2 k_{x3})] -\cos[d (- 2 k_{x1}+k_{x2})])  -(\chi_2 \chi_5 -  \chi_1 \chi_6)
		(\sin[d(- 2 k_{x1}+k_{x2}+2 k_{x3})]\nonumber\\
		& -\sin[d(- 2 k_{x1}+k_{x2})])  +(\chi_3 \chi_5+ \chi_4 \chi_6) (\cos(-2 d k_{x1})-\cos[2 d(-k_{x1}+k_{x2}+k_{x3})])\nonumber\\
		& -(\chi_4 \chi_5- \chi_3 \chi_6)(\sin(-2 d k_{x1})-\sin[2 d(-k_{x1}+k_{x2}+k_{x3})]) \nonumber
		\\
		b_2 =& (\chi_1 \chi_7+  \chi_2 \chi_8) (\sin[2d (k_{x3}-k_{x1})]-\sin[2d (k_{x2}-k_{x1})])  
		+(\chi_2 \chi_7-\chi_1 \chi_8) (\cos[2 d(k_{x3}-k_{x1})] \nonumber \\
		&-\cos[2 d(k_{x2}-k_{x1})]) 
		+(\chi_3 \chi_7+ \chi_4 \chi_8 ) (\sin[d(- 2 k_{x1}+k_{x2})]-\sin[d (- 2 k_{x1}+k_{x2}+2 k_{x3})])\nonumber \\
		& +(\chi_4 \chi_7- \chi_3 \chi_8 )(\cos[ d(- 2 k_{x1}+k_{x2})]-\cos[ d(- 2 k_{x1}+k_{x2}+2 k_{x3})])  
		+(\chi_1 \chi_5+  \chi_2 \chi_6) \\
		&(\sin[d(- 2 k_{x1}+k_{x2}+2 k_{x3})]-\sin[d(- 2 k_{x1}+k_{x2})])+(\chi_2 \chi_5-  \chi_1 \chi_6)(\cos[d(- 2 k_{x1}+k_{x2}+2 k_{x3})]\nonumber \\
		&-\cos[d (- 2 k_{x1}+k_{x2})])+(\chi_3 \chi_5+ \chi_4 \chi_6) (\sin(-2 d k_{x1})-\sin[2 d(-k_{x1}+k_{x2}+k_{x3})]) 
		\nonumber \\
		&+(\chi_4 \chi_5- \chi_3 \chi_6) (\cos(-2 d k_{x1})-\cos[2 d(-k_{x1}+k_{x2}+k_{x3})]) \nonumber
		\\  
		a_3 =&  -(\chi_1^2+\chi_2^2) \cos(2 d k_{x3} )-(\chi_3^2+\chi_4^2) \cos[2 d(k_{x2}+k_{x3})] +2 ( \chi_1 \chi_3+ \chi_2 \chi_4)\cos[d(k_{x2}+2 k_{x3})] \nonumber \\
		& +(\chi_5^2+\chi_6^2) +(\chi_7^2+\chi_8^2) \cos (2 d k_{x2})+ 2 ( \chi_5 \chi_7+  \chi_6 \chi_8) \cos d(d k_{x2}) 
		\\  
		b_3 =&-(\chi_1^2+\chi_2^2) \sin(2 d k_{x3} )-(\chi_3^2+\chi_4^2) \sin[2 d(k_{x2}+k_{x3})]+2 ( \chi_1 \chi_3+ \chi_2 \chi_4) \sin[d(k_{x2}+2 k_{x3})]\nonumber \\
		&+(\chi_7^2+\chi_8^2) \sin (2 d k_{x2})+2 ( \chi_5 \chi_7+  \chi_6 \chi_8) \sin d(d k_{x2}) 
	\end{align}
	\begin{align}  
		&\chi_1=-\frac{k_{x1} k_{x2}-k_y^2}{A_1 A_2}+\frac{k_{x1} k_{x3}-k_y^2}{A_1 A_3}+\frac{k_{x2} k_{x3}+k_y^2}{A_2 A_3}-\frac{k_{x2}^2+k_y^2}{A_2^2}
		\\
		& \chi_2= \tau k_y \left(\frac{k_{x1}+k_{x2}}{A_1 A_2}-\frac{k_{x1}+k_{x3}}{A_1 A_3}+\frac{k_{x3}-k_{x2}}{A_2 A_3}\right)
		\\
		& \chi_3= \frac{k_{x1} k_{x2}+k_y^2}{A_1 A_2}+\frac{k_{x1} k_{x3}-k_y^2}{A_1 A_3}-\frac{k_{x2} k_{x3}-k_y^2}{A_2 A_3}-\frac{k_{x2}^2+k_y^2}{A_2^2}
		\\
		& \chi_4= \tau k_y \left(\frac{k_{x1}-k_{x2}}{A_1 A_2}-\frac{k_{x1}+k_{x3}}{A_1 A_3}+\frac{k_{x2}+k_{x3} }{A_2 A_3 }\right)
		\\
		& \chi_5= \frac{k_{x1} k_{x2}-k_y^2}{A_1 A_2}+\frac{k_{x1} k_{x3}+k_y^2}{A_1 A_3}+\frac{k_{x2} k_{x3}-k_y^2}{A_2 A_3}+\frac{k_{x2}^2+k_y^2}{A_2^2}
		\\
		&
		\chi_6= \tau k_y \left(\frac{k_{x1}+k_{x2} }{A_1 A_2}-\frac{k_{x1}-k_{x3} }{A_1 A_3}-\frac{k_{x2}+k_{x3}}{A_2 A_3}\right)
		\\
		&
		\chi_7= \frac{k_{x1} k_{x2}+k_y^2}{A_1 A_2}-\frac{k_{x1} k_{x3}+k_y^2}{A_1 A_3}+\frac{k_{x2} k_{x3}+k_y^2}{A_2 A_3}-\frac{k_{x2}^2+k_y^2}{A_2^2}
		\\
		&
		\chi_8= \tau k_y \left(-\frac{k_{x1}-k_{x2}}{A_1 A_2}+\frac{k_{x1}-k_{x3}}{A_1 A_3}-\frac{k_{x2}-k_{x3} }{A_2 A_3}\right)
		\\
		&
		A_j=E+\beta \lambda_{ex}^B +\Delta-V_j.
	\end{align}
\end{widetext}

\end{document}